\documentclass[conference, 10pt]{IEEEtran}

\usepackage[T1]{fontenc}
\usepackage{graphicx}
\usepackage[cmex10]{amsmath}
\usepackage{array}
\usepackage{fixltx2e}
\usepackage{url}
\usepackage{amsfonts}
\usepackage{lmodern}
\usepackage{algpseudocode}
\usepackage{algorithmicx}
\usepackage{algorithm}
\usepackage{xfrac}
\usepackage{caption}
\usepackage{subcaption}
\usepackage{float}
\usepackage{multirow}
\usepackage{balance}

\DeclareMathOperator{\prox}{prox}

\DeclareMathOperator{\soft}{\op{S}}

\DeclareMathOperator{\proj}{\op{P}}

\DeclareMathOperator*{\minimise}{minimise}

\newcommand{\ds}{\displaystyle}

\newcommand{\bs}{\boldsymbol}
\newcommand{\bm}[1]{\boldsymbol{\mathrm #1}}
\newcommand{\bb}{\mathbb}
\newcommand{\mc}{\mathcal}
\newcommand{\diff}{\mathrm{d}\hspace{-0.1ex}}
\newcommand{\acp}[1]{(\uppercase{#1})}
\newcommand{\ac}[1]{\uppercase{#1}}
\newcommand{\alg}[1]{\uppercase{#1}}
\newcommand{\op}[1]{\boldsymbol{\mc{#1}}}

\algblock[Block]{Block}{EndBlock}
\algblockdefx[Block]{Block}{EndBlock}%
[1]{{#1}}%
[1]{{#1}}

\newcommand{\Given}[1]{\State{\bf given} {#1}}

\newcommand{\RepeatFor}[1]{\Repeat {\bf~for} {#1}}

\newcommand{\ParForD}[2]{\Block{{#1} {\bf distribute} {#2} {\bf and do in parallel}}}
\newcommand{\EndParForD}[1]{\EndBlock{\bf end and gather} {#1}}

\IEEEoverridecommandlockouts

\title{A Randomised Primal-Dual Algorithm for Distributed Radio-Interferometric Imaging}

\author{
\IEEEauthorblockN{
Alexandru Onose\IEEEauthorrefmark{1},
Rafael E. Carrillo\IEEEauthorrefmark{2},
Jason D. McEwen\IEEEauthorrefmark{3}
and Yves Wiaux\IEEEauthorrefmark{1}}

\thanks{This work was supported by the UK Engineering and Physical Sciences Research Council (EPSRC, grants EP/M011089/1 and EP/M008843/1) and by the UK Science and Technology Facilities Council (STFC, grant ST/M00113X/1), as well as by the Swiss National Science Foundation (SNSF) under grant 200020-146594.}

\IEEEauthorblockA{~\\ \IEEEauthorrefmark{1}Institute of Sensors, Signals and Systems, Heriot-Watt University, Edinburgh EH14 4AS, United Kingdom}
\IEEEauthorblockA{\IEEEauthorrefmark{2}Signal Processing Laboratory (LTS5), Ecole Polytechnique F\'{e}d\'{e}rale de Lausanne, Lausanne CH-1015, Switzerland}
\IEEEauthorblockA{\IEEEauthorrefmark{3}Mullard Space Science Laboratory,  University College London, Surrey RH5 6NT, United Kingdom}
}

\begin{document}
\maketitle

\begin{abstract}
Next generation radio telescopes, like the Square Kilometre Array, will acquire an unprecedented amount of data for radio astronomy.
The development of fast, parallelisable or distributed algorithms for handling such large-scale data sets is of prime importance.
Motivated by this, we investigate herein a convex optimisation algorithmic structure, based on primal-dual forward-backward iterations, for solving the radio interferometric imaging problem.
It can encompass any convex prior of interest.
It allows for the distributed processing of the measured data and introduces further flexibility by employing a probabilistic approach for the selection of the data blocks used at a given iteration.
We study the reconstruction performance with respect to the data distribution and we propose the use of nonuniform probabilities for the randomised updates.
Our simulations show the feasibility of the randomisation given a limited computing infrastructure as well as important computational advantages when compared to state-of-the-art algorithmic structures.
\end{abstract}


\section{Introduction}

Radio interferometry \acp{ri} is a technique that, by measuring signal correlations between geographically separated antennas, is able to greatly improve the sensitivity and angular resolution with which we can observe radio emissions.
Extensively studied, it provides valuable data driving many research directions in cosmology, astronomy, and astrophysics  \cite{thompson01}.
It has greatly contributed to the advancement of our understanding of the universe.
In the future, the Square Kilometre Array, is expected to acquire massive amounts of data which will be used to reconstruct giga-pixel sized images, two orders of magnitude improvement over current instruments \cite{Dewdney2009}.
Under these big-data challenges, the methods solving the image reconstruction problem need be computationally efficient and to scale well in order to work with a huge number of measurements.

Recently, advances in sparse modelling and convex optimisation allowed the development of efficient algorithms able to outperform the standard \ac{ri} methods in terms of both reconstruction quality and computational cost \cite{Carrillo2014}.
Modern optimisation techniques, like the proximal splitting and the primal-dual \acp{pd} methods, are of particular importance since they decompose the main minimisation problem into multiple, simpler, subproblems for different data blocks or priors that are solved in parallel \cite{combettes11, Komodakis2015}.

Herein, we study a randomised algorithmic structure, based on the \ac{pd} framework, for solving a sparse regularization problem in the context of RI imaging \cite{carrillo12}.
We review the algorithmic structure from \cite{Onose2016} and present a new strategy for the choice of probabilities that significantly improves the convergence speed.
The algorithm achieves the full splitting of the functions and operators used and has a highly parallelisable structure.
The computational cost is much lower than that of the simultaneous direction method of multipliers \acp{sdmm} \cite{Carrillo2014}, previously proposed for a similar problem.
Through randomisation, the method achieves great flexibility, in terms of computational burden per iteration and memory load, at the cost of requiring more iterations to converge.
An adaptive nonuniform choice for the probabilities provides an excellent tradeoff between complexity per iteration and convergence speed.

The contents of the paper are as follows.
In Section~\ref{sec-ri} we present the RI imaging problem and describe the current image reconstruction techniques. Section~\ref{sec-pd} introduces the optimisation problem and details the \ac{pd} algorithm. We also discuss implementation and computational complexity details.
Numerical experiments, to assess the algorithm performance, are reported in Section~\ref{sec-results}.

\section{Radio-interferometric imaging}
\label{sec-ri}
Radio-interferometric data, the visibilities, are signals produced by an array of antenna pairs measuring the intensity of the sky brightness.
They are related to the baseline components defined by the relative position of each pair of telescopes.
These components, usually denoted $u$, $v$, and $w$, identify the orthogonal plane $\bs{u}=(u,v)$ and the direction of observation $w$.
The measured sky brightness $x$ is expressed in the same coordinate system, with components $l$, $m$, and $n$.
For non-polarised monochromatic \ac{ri} imaging, each visibility $y(\bs{u})$ is generally modelled by
\begin{equation}
	y(\bs{u}) = \int D(\bs{l},\bs{u}) x(\bs{l}) e^{-2 i \pi \bs{u} \cdot \bs{l}} \diff^2 \bs{l},
	\label{measurement-eq}
\end{equation}
with $D (\bs{l}, \bs{u}) = \tfrac{1}{n(\bs{l})} \bar{D} (\bs{l},\bs{u})$ quantifying the different direction dependant effects (DDEs) that affect the measurements, $\bs{l}=(l,m)$, and $n(\bs{l})=\sqrt{1-l^2-m^2}$, $l^2+m^2 \leq 1$.
This defines an inverse problem with the sky brightness $x$ to be recovered from the measurements $y$.
By discretising (\ref{measurement-eq}), we can define a linear measurement model as 
\begin{equation}
	\bs{y} = \bm{\Phi} \bs{x} + \bs{n},
	\label{inverse-problem}
\end{equation}
where the measurement operator $\bm{\Phi} \in \bb{C}^{M \times N}$ maps the image domain to the $u$--$v$ visibility space and the measured visibilities $\bs{y}$ are corrupted by some additive noise $\bs{n}$.

Limitations in the sampling scheme, due to the physical position of the telescopes, make (\ref{inverse-problem}) an ill-posed inverse problem. The large dimensionality of the problem, $M \gg N$, also presents an important challenge due to high memory and computational requirements. 
Thus, a fast implementation of all operators involved in the image reconstruction is essential.
The measurement operator,
\begin{equation}
	\bm{\Phi} = \bm{G} \bm{F} \bm{Z},
\end{equation}
is formed by a $k$-oversampled Fourier operator premultiplied by a sparse matrix $\bm{G} \in \bb{C}^{M \times kN}$ that models the compact support \ac{dde}s as a convolution in the Fourier domain and performs a local interpolation of nearby uniformly sampled Fourier coefficients $\bm{F} \bm{Z} \bs{x}$ \cite{Fessler2003}.
The matrix $\bm{Z} \in \bb{R}^{kN \times N}$ performs the oversampling and pre-compensates for the interpolation.

The standard, most widely used imaging methods belong to the \alg{clean} family.
They perform a greedy non-linear deconvolution by iteratively searching for atoms to be included in the solution \cite{thompson01}.
The method is related to the matching pursuit algorithm \cite{mallat93}, implicitly introducing sparsity in the image through the greedy selection.
Convex optimisation algorithms, using sparsity-aware models, have also been proposed under the compressed sensing \acp{cs} theoretical framework.
Their reconstruction quality is generally superior if compared to \alg{clean} \cite{carrillo12}.

Following the \ac{cs} approach, the underlying signal is considered to have a sparse representation in a dictionary $\bm{\Psi}$, $\bm{\Psi}^\dagger \bs{x}$ containing only a few nonzero elements.
Analysis-based approaches recover the signal solving the constrained minimisation problem
\begin{equation}
	\minimise_{\bs{x}} \| \bm{\Psi}^\dagger \bs{x} \|_1 \quad \rm{subject~to} \quad \|\bs{y} - \bm{\Phi} \bs{x}\|_2 \leq \epsilon.
	\label{analysis-l1-problem}
\end{equation}
Under this formulation, the sparsity averaging reweighed analysis \acp{sara} \cite{carrillo12} shows superior reconstruction quality when compared to the \ac{clean} methods.
A synthesis approach has also been proposed \cite{wiaux09, mcewen11a}.

Current algorithmic solvers are not specifically designed for large-scale data sets.
The \ac{sdmm} algorithm, recently proposed for \ac{ri} imaging \cite{Carrillo2014}, allows for an efficient large-scale distributed processing.
It however requires the solution to a linear system of equations at each iteration, which is expensive to compute when large images have to be recovered.
The \ac{pd} algorithm herein achieves the full splitting of both operators and functions used to define the minimisation task.
The level of parallelism and distribution is similar to that of \ac{sdmm}.
No matrix inversion or solution to linear systems of equations is required however.
The randomised updates further increase the scalability.

\section{Primal-dual forward-backward algorithm}
\label{sec-pd}


The \ac{cs} paradigm allows the redefinition of the inverse problem (\ref{inverse-problem}) by adding the additional constraint that the image is sparse in an over-complete dictionary $\bm{\Psi}$.
Since $\bs{x}$ represents an intensity image, a positivity requirement is also added to the solution.
Including all these constraints and using the \ac{cs} analysis formulation, we define the reconstruction task as the convex minimisation problem 
\begin{equation}
	\minimise_{\bs{x}} f(\bs{x}) + \gamma l(\bm{\Psi}^\dagger \bs{x}) + \sum_{i=1}^{d} h_i(\bm{\Phi}_i\bs{x})
	\label{basic-min-problem}
\end{equation}
with the functions involved, defined as
\begin{equation}
	\begin{aligned}
		f(\bs{z})  & = \iota_{\mc{D}} (\bs{z}), \mc{D} = \bb{R}^N_+, \quad l(\bs{z})  = \| \bs{z} \|_1, \\
		h_i(\bs{z}) & = \iota_{\mc{B}_i}  (\bs{z}), \mc{B}_i = \{ \bs{z} \in \bb{C}^{M_i} : \| \bs{z} - \bs{y}_i \|_2 \leq \epsilon_i \},
	\end{aligned}
	\label{basic-min-problem-function-def}
\end{equation}
being proper, lower semicontinuous, and convex. The operator $^\dagger$ denotes the adjoint of the linear operator.
We include the constraints through the indicator function,
\begin{equation}
	\iota_{\mc{C}} (\bs{z}) \overset{\Delta}{=} \left\{ \begin{aligned}
					0 & \qquad \bs{z} \in \mc{C} \\
					+\infty & \qquad \bs{z} \notin \mc{C},
				   \end{aligned} \right.
	\label{indicator-function}
\end{equation}
of the convex sets $\mc{D}$ and $\mc{B}_i$ defining the feasibility regions.

For an efficient parallel implementation, the data are split into multiple blocks 
\begin{equation}
    		\bs{y} = \begin{bmatrix}
                		\bs{y}_1 \\
                		\vdots \\
                		\bs{y}_{d}
                	\end{bmatrix}, \qquad
                	\bm{\Phi} = \begin{bmatrix}
                		\bm{\Phi}_1 \\
                		\vdots \\
                		\bm{\Phi}_{d}
                	\end{bmatrix}
                	= \begin{bmatrix}
                		\bm{G}_1 \bm{M}_1\\
                		\vdots \\
                		\bm{G}_{d} \bm{M}_{d} 
                	\end{bmatrix}  \bm{F}\bm{Z}.
        	\label{data-split}
\end{equation}
The computations involving the compact support kernels modelled through $\bm{G}_i \in \bb{C}^{M_i \times kN_i}$ only require parts of the discrete Fourier plane, selected through the mask matrices $\bm{M}_i \in \bb{R}^{kN_i \times kN}$.
The resulting blocks are processed in parallel, each one dealing with a limited number of discrete frequency points under the assumption that the blocks $\bs{y}_i$ cover a compact region in the $u$--$v$ space.
The sparsity operator $\bm{\Psi} \in \bb{C}^{N \times bN}$ is chosen as a collection of $b$ bases \cite{Carrillo2014}.
We use herein the \ac{sara} wavelet bases \cite{carrillo12} but problem (\ref{basic-min-problem-function-def}) is not restricted to them.

The minimisation defined in (\ref{basic-min-problem}), referred to as the primal problem, accepts a dual problem,
\begin{equation}
		\minimise_{\bs{s}, \bs{v}_i} f^* \Bigg(-\bm{\Psi}\bs{s} - \sum_{i=1}^{d} \bm{\Phi}^\dagger_i \bs{v}_i \Bigg)  + \frac{1}{\gamma} l^*(\bs{s}) + \sum_{i=1}^{d} h_i^*(\bs{v}_i),
	\label{split-min-dual-problem}
\end{equation}
where $^*$ denotes the Legendre-Fenchel conjugate function, defined for any function $g$ as $g^*(\bs{v}) = \sup_{\bs{z}} \bs{z}^\dagger \bs{v} - g(\bs{z})$.
The algorithm converges to a Kuhn-Tucker point $(\hat{\bs{x}}, \hat{\bs{s}}, \hat{\bs{v}}_i)$ with $\hat{\bs{x}}$ the solution to the primal problem and $(\hat{\bs{s}}, \hat{\bs{v}}_i)$ the solution to the dual one.
To deal with the non smooth functions, forward-backward iterations are used \cite{Komodakis2015}.
They combine a forward, gradient-like step with the application of the proximity operator for the non smooth functions, which implicitly performs a sub-gradient-like backward step.
A \ac{pd} approach with forward-backward iterations for solving (\ref{basic-min-problem}) and (\ref{split-min-dual-problem}) produces a highly paralelisable, fast algorithm \cite{Pesquet2014}. Its structure is presented in Algorithm \ref{alg-primal-dual}.

\begin{algorithm}[t]
\caption{Randomised forward-backward \ac{pd}.}
\label{alg-primal-dual}

\begin{algorithmic}[1]
\small

\Given{$\bs{x}^{(0)}, \tilde{\bs{x}}^{(0)}, \bs{s}^{(0)}, \bs{v}_j^{(0)}, \tilde{\bs{v}}^{(0)}_j, \kappa, \tau, \sigma, \varsigma$}

\RepeatFor{$t=1,\ldots$}
\State {\bf generate set} $\mc{A} \subset \{1,\ldots, d\}$
\State $\ds \tilde{\bs{b}}^{(t)} = \bm{F}\bm{Z} \tilde{\bs{x}}^{(t-1)}$
\State $\forall i \in \mc{A}$ {\bf set} $\ds \bs{b}_i^{(t)} = \bm{M}_i \tilde{\bs{b}}^{(t)}$

\Block{\bf run concurrently the 3 blocks}
\ParForD {$\forall i \in \mc{A}$}{$\bs{b}_i^{(t)}$}
	\State $\ds \bs{v}_i^{(t)} \!= \! \bs{v}_i^{(t-1)} + \bm{G}_i \bs{b}^{(t)}_i - \proj_{\mc{B}_i} \Big( \bs{v}_i^{(t-1)} + \bm{G}_i \bs{b}^{(t)}_i \Big)$
	\State $\ds \tilde{\bs{v}}^{(t)}_i \!= \! \bm{G}_i^\dagger \bs{v}^{(t)}_i$
\EndParForD{$\tilde{\bs{v}}^{(t)}_i$}
\State $\forall i \in \{1, \ldots d\} \setminus \mc{A}$ {\bf set} $\ds \bs{v}^{(t)}_i \!= \bs{v}^{(t-1)}_i$ {\bf and} $\ds \tilde{\bs{v}}^{(t)}_i \!= \tilde{\bs{v}}^{(t-1)}_i$

\Block{\bf do}
	\State $\ds \bs{s}^{(t)} \!= \!\bs{s}^{(t-1)}\! + \bm{\Psi}^\dagger  \tilde{\bs{x}}^{(t-1)} \!\! - \soft_{\!\kappa \|\bm{\Psi}\|_{\rm{S}}} \!\!\Big(\! \bs{s}^{(t-1)} \!+\! \bm{\Psi}^\dagger  \tilde{\bs{x}}^{(t-1)} \!\!\Big)$
	\State $\ds \tilde{\bs{s}}^{(t)} \!= \! \bm{\Psi} \bs{s}^{(t)}$
\EndBlock{\bf end}

\EndBlock{\bf end}

\State $\ds \bs{x}^{(t)} \! = \proj_{\mc{C}} \! \bigg(\! \bs{x}^{(t-1)}  - \tau  \Big(\sigma \tilde{\bs{s}}^{(t)} + \varsigma \bm{Z}^\dagger \bm{F}^\dagger \sum_{i=1}^{d} \bm{M}_i^\dagger \tilde{\bs{v}}_i^{(t)} \Big)\!\!\bigg)$
\State $\tilde{\bs{x}}^{(t)}=2\bs{x}^{(t)} - \bs{x}^{(t-1)}$
\Until {\bf convergence}

\end{algorithmic}
\end{algorithm}

All the dual variables are updated in parallel in steps $6$ to $16$ with forward-backward iterations.
Data fidelity is enforced by constraining the residuals to belong to the $\ell_2$ balls $\mc{B}_i$ through the application of the proximity operator of the conjugate functions $h_i^*$.
The Moreau decomposition $\bs{z} = \prox_{\alpha g} (\bs{z}) + \alpha \prox_{\alpha^{-1}g^*}(\alpha^{-1}\bs{z})$, $0 < \alpha < \infty$,
is used to replace the proximity operator of the conjugate $h_i^*$, with that of the original function $h_i$. This accepts a closed form solution as the projection
\begin{equation}
	\proj_{\mc{B}_i}(\bs{z})  \overset{\Delta}{=} \left\{ 
	\begin{aligned}
		\epsilon_i \frac{\bs{z} - \bs{y}_i}{\|\bs{z} - \bs{y}_i\|_2} + \bs{y_i} & \quad \|\bs{z} - \bs{y}_i\|_2 > \epsilon_i\\
		\bs{z} \qquad \qquad & \quad \|\bs{z} - \bs{y}_i\|_2 \leq \epsilon_i
	\end{aligned}\right.
	\label{proj-L2}
\end{equation}
onto the ball $\mc{B}_i$.
The operations are presented in Algorithm \ref{alg-primal-dual}, step $8$.
All computations can be distributed, the communication of the partial Fourier information $\bs{b}_i^{(t)} \in \bb{C}^{kN_i}$, computed locally in steps $4$ and $5$, being feasible


The algorithm benefits from a very interesting randomisation functionality.
It performs the update only for an active subset $\mc{A}$ of the data blocks.
This active set $\mc{A}$ is generated in step $3$ assuming that each data block $\bs{y}_i$ has a probability $p_i$ of being used at the current iteration.
If a block remains inactive, the associated dual variables remain unchanged (step $11$).
This lowers the computational and memory cost per iteration, which can be important given processing infrastructure constraints.

For the sparsity prior, a similar approach, using the Moreau decomposition, is detailed in step $13$.
This involves the application of the proximity operator to the sparsity prior function $l$, which is a soft-thresholding operator, defined component wise for a threshold $\alpha$ as
\begin{equation}
	\Big( \soft_{\alpha}(\bs{z}) \Big)_k \overset{\Delta}{=} \left\{ 
	\begin{array}{cl}
		\ds \frac{z_k\{ | z_k | - \alpha \}_{+}}{| z_k |} & \quad | z_k | > 0\\
		\ds 0 & \quad | z_k | = 0\\
	\end{array}\right. \quad \forall k.
	\label{prox-L1}
\end{equation}
Since the minimisation problem is defined with the free parameter $\gamma$, we replace the resulting algorithmic soft-threshold size $\gamma\sigma^{-1}$ with $\kappa \|\bm{\Psi}\|_{\rm{S}}$, where $\|\bm{\Psi}\|_{\rm{S}}$ represents the spectral norm of the operator.
This new scale-free parameter $\kappa$ can be seen as a normalised soft-thresholding value and is independent to the operator $\bm{\Psi}$.

The computation of the primal variable from step $17$ relies on the same forward-backward approach.
It consists of a gradient-like update using the dual information from $\tilde{\bs{v}}_i$ and $\tilde{\bs{s}}$, computed in parallel in steps $9$ and $14$.
This is followed by the application of the proximity operator to the function $f^*$.
Using the Moreau decomposition, the operation resolves to a projection onto the positive orthant, component wise defined as
\begin{equation}
	\Big( \proj_{\mc{C}}(\bs{z}) \Big)_k  \overset{\Delta}{=} \left\{ 
	\begin{array}{cl}
		\Re(z_k) & \qquad \Re(z_k) > 0 \\
		0 & \qquad \Re(z_k) \leq 0
	\end{array}\right. \quad \forall k.
	\label{proj-plus}
\end{equation}
The communication of the data fidelity update $\tilde{\bs{v}}_i$ is feasible on a distributed system since each block updates only parts of the frequency information.

\subsection{Implementation details}

For an efficient parallel processing, the computations are preformed on a central meta-node and a collection of distributed processing nodes.
The central node updates the solution image $\bs{x}^{(t)}$ and computes its Fourier transform to be distributed to the other nodes.
It also performs the sparsity prior computations.
The computations involving the data fidelity terms are distributed and processed while the central node updates the sparsity prior dual variable.
The data distribution communication cost is low with only $kN_i$ coefficients to be transferred to each node $i$.

The application of the oversampled Fourier operators $\bm{F}$ and $\bm{F}^\dagger$ scales as $\mc{O}\left( kN \log kN \right)$.
The sparsity prior computations have a complexity $\mc{O}(N)$ for compact support wavelets.
For the data nodes, the data are split into similarly sized blocks spanning a compact frequency range in the $u$--$v$ space.
Since the interpolation kernels \cite{Fessler2003} and DDEs \cite{wolz13} have compact support, each block requires only $kN_i$ discrete Fourier coefficients linked to the blocks' frequency range. An overlap linked to the interpolation kernel and DDE support size must exist between adjacent blocks.
The matrix $\bm{G}_i$ is sparse, with sparsity $s$, the computational complexity for the data fidelity terms per block being $\mc{O}(sM_i kN_i)$. 
The number of blocks $d$ is linked to the available nodes $n$.
A randomised approach lowers the required number of nodes since only part of the data blocks are processed per iteration.
This is fundamental when dealing with limited resources, number of nodes and memory, and makes the algorithm extremely scalable.

\subsection{Algorithm convergence}
The convergence of Algorithm \ref{alg-primal-dual} is achieved if the parameters $\tau$, $\sigma$ and $\varsigma$ satisfy $\tau \left ( \sigma  \| \bm{\Psi}^\dagger  \|_{\rm{S}}^2 +  \varsigma  \| \bm{\Phi}  \|_{\rm{S}}^2 \right )<1$ \cite{Pesquet2014}.
The parameter $\kappa$ needs to be positive since $\gamma > 0$.
The probabilities $p_i$ are nonzero, fixed such that an entry of $\mc{A}$ is chosen independently at each iteration.

\section{Adapting the probabilities}
We quantify the signal energy present in the residual $\bm{\Phi}_i\bs{x}^{(t)} - \bs{y}_i$, for block $i$,  
using the relative distance of the residual to the ball $\mc{B}_i$,
\begin{equation}
c_i = \frac{\max(\|\bm{\Phi}_i\bs{x}^{(t)} - \bs{y}_i\|_2 - \epsilon_i, 0)^2}{\epsilon_i^2}.
\end{equation}
For a fixed number of nodes $n$ with limited available memory, splitting the data into multiple blocks $d>n$ that fit into memory and using the randomised algorithm may be the only feasible approach.
We propose a nonuniform adaptive choice for the probabilities $p_i$ proportional to the quantities $c_i$ under the constraint that on average we use $n$ nodes and as such $\sum_{i=1}^{d} p_i = n$.
This results in setting
\begin{equation}
p_i = \check{p} + \min \left(\Big(n - d\check{p}\Big) \frac{c_i}{\sum_{i=1}^d c_i} + a, \hat{p} - \check{p}\right),
\label{prob-choice}
\end{equation}
where we denote by $\check{p}$ and $\hat{p}$ a minimum and a maximum value for the probabilities, respectively.
They need to satisfy $\check{p} \leq n/d$ and  $\check{p} \leq \hat{p}$. If $\sum_{i=1}^d c_i=0$, we set uniform probabilities $p_i = n/d, \forall i$.
The parameter $a$ serves to rebalance the choice of probabilities and is nonzero only if some values $p_i$ achieve saturation $p_i = \hat{p}$.
In such a case, it is increased until the constraint $\sum_{i=1}^{d} p_i = n$ is satisfied.

\section{Simulation results}
\label{sec-results}

\begin{figure*}
	\centering
	\includegraphics[trim={30px 20px 30px 20px}, clip, height=3.8cm]{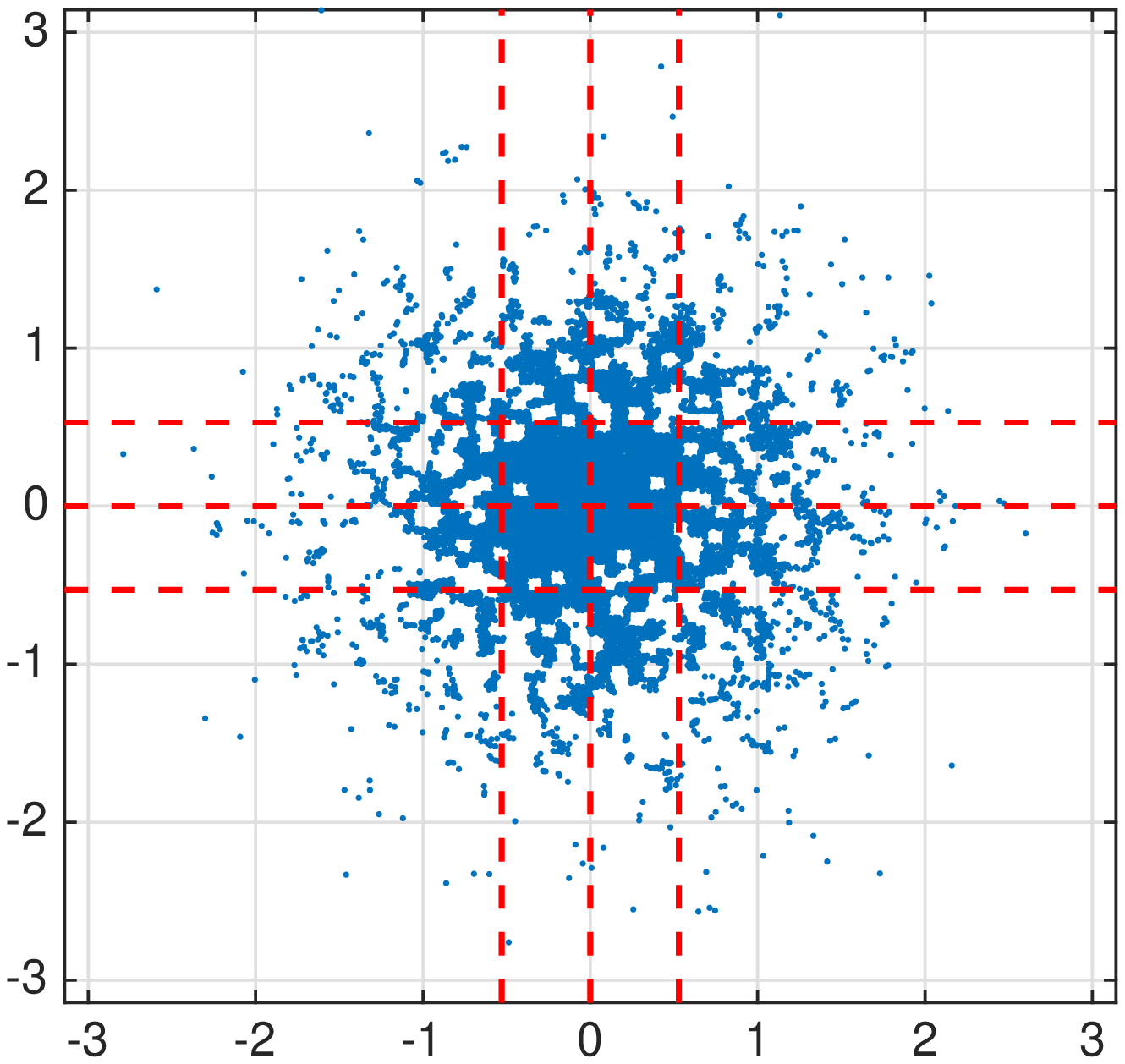}
	\hspace{20pt}
  	\includegraphics[trim={0px 0px 0px 0px}, clip, height=3.8cm]{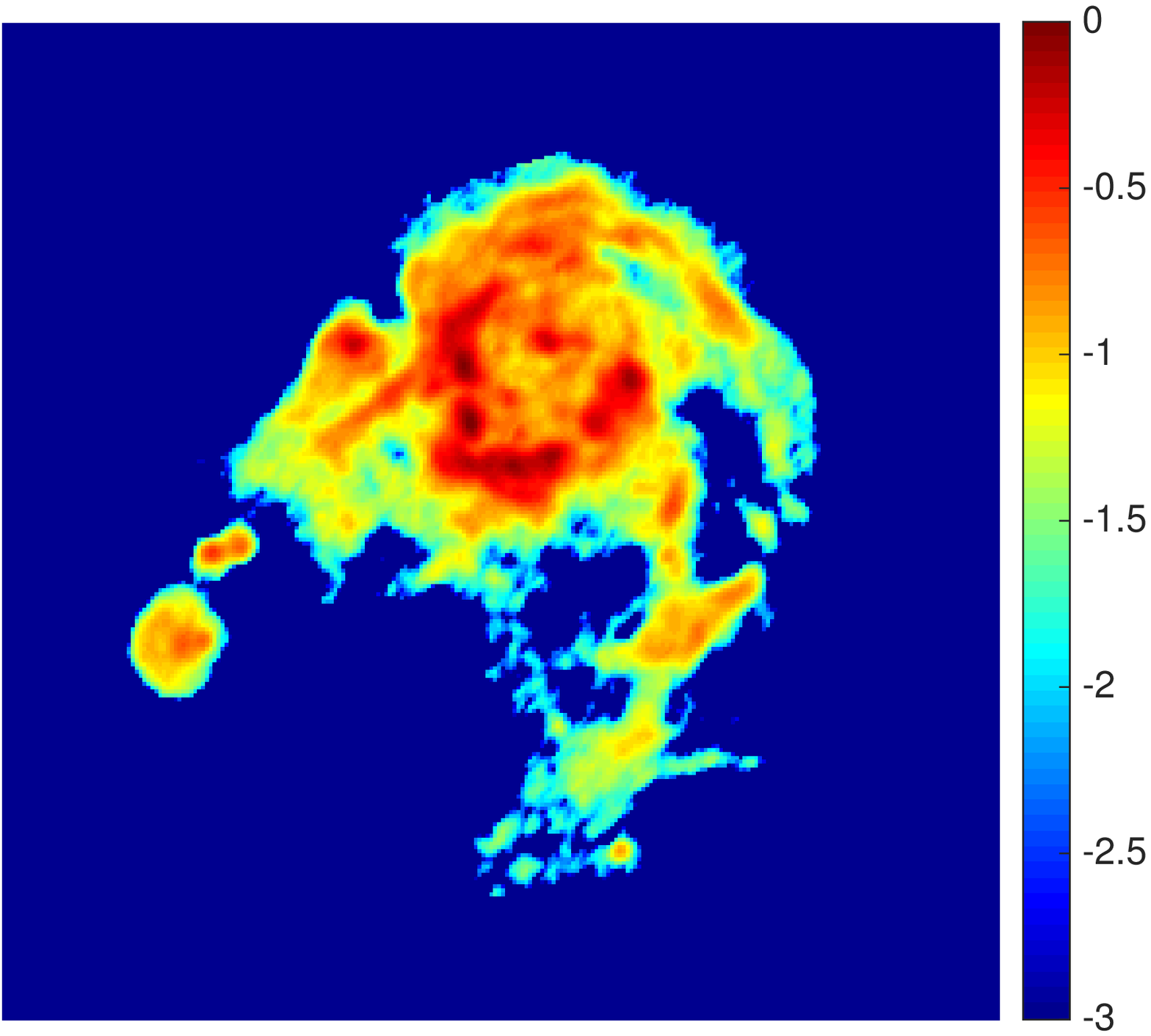}
	\hspace{20pt}
	\includegraphics[trim={0px 0px 0px 0px}, clip, height=3.8cm]{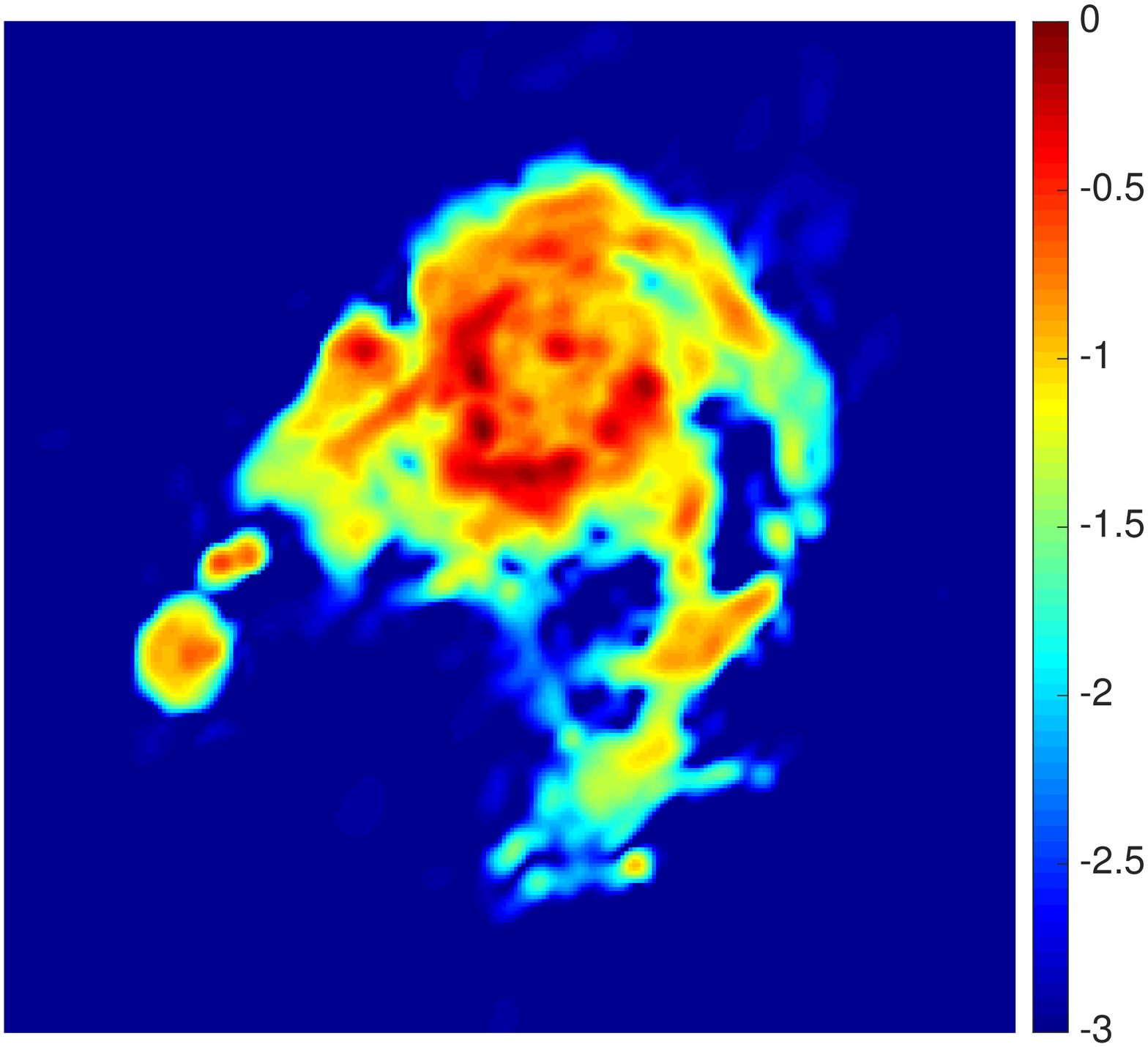}
	\vspace{-2px}
	\caption{(left) Randomly generated $u$--$v$ coverage with $M=16384$ visibilities. The visibilities are split into $16$ blocks marked by red dashed lines. (centre) The $256 \times 256$ M31 galaxy test image in log scale. (right) The log scale reconstructed image produced by PD-R using adaptive probabilities for $\sum_{i=1}^d p_i=4$, $\check{p} = 0.05$ and $\hat{p}=0.5$. The SNR is $26.6$ dB.
}
	\label{fig-images-and-coverage}
	\vspace{-10px}
\end{figure*}

We study the behaviour of \ac{pd-r} for the reconstruction of the $256 \times 256$ image of the H\textsc{ii} region of the M31 galaxy from simulated data. The image is presented in Fig.~\ref{fig-images-and-coverage}.
The solution to (\ref{basic-min-problem}) was reported to produce superior reconstruction quality when compared to the state-of-the-art \alg{clean} methods \cite{Carrillo2014, carrillo12}.
We use \ac{sdmm}, solving (\ref{basic-min-problem}), as a proxy for the performance \cite{Carrillo2014}.
The algorithms are as follows: \ac{pd-r}, the algorithm from Algorithm \ref{alg-primal-dual}, solving (\ref{basic-min-problem}); \ac{sdmm}, the method from \cite{Carrillo2014} solving (\ref{basic-min-problem}).

We generate the $u$--$v$ coverage randomly through Gaussian sampling, with $0$ mean and variance of $0.25$ of the maximum frequency.
To simulate the incomplete sampling typically available in \ac{ri}, we introduce holes predominantly in the high frequency data using an inverse Gaussian profile.
We use $M=5N$ measurements, split into $d = 16$ equal size blocks.
An example with normalised frequencies is presented in Fig.~\ref{fig-images-and-coverage}.
The measurement data are corrupted by complex Gaussian noise with $0$ mean producing a signal to noise level of $20~\rm{dB}$.
The bounds $\epsilon_i$, estimated based on the noise level, are set to the $0.99$ percentile of the $\chi^2$ distribution with $2M_i$ degrees of freedom, which models the residual norm.
We assume no DDEs and we use $8 \times 8$ interpolation kernels \cite{Fessler2003}.
An oversampling ratio is $k=4$.

The parameters are $\sigma = \tfrac{1}{\| \bm{\Psi} \|_{\rm{S}}^2}$, $\varsigma = \tfrac{1}{\| \bm{\Phi} \|_{\rm{S}}^2}$ and $\tau=0.49$.
All operator norms are computed a priori using the power method.
The normalised soft-thresholding is set $\kappa = 10^{-3}$. This was observed to produce the fastest convergence speed.
The parameters for \ac{sdmm} are as suggested by \cite{Carrillo2014}.
The performance is measured by the signal to noise ratio
\begin{equation}
	{\rm SNR} = 20 \log_{10} \left( \frac{\|\bs{x}\|_2}{\|\bs{x} - \hat{\bs{x}}\|_2}\right),
\end{equation}
averaged over $10$ tests with different noise realisations.

\begin{figure}
	\centering
	\includegraphics[trim={0px 0px 0px 0px}, clip, width=0.9\linewidth]{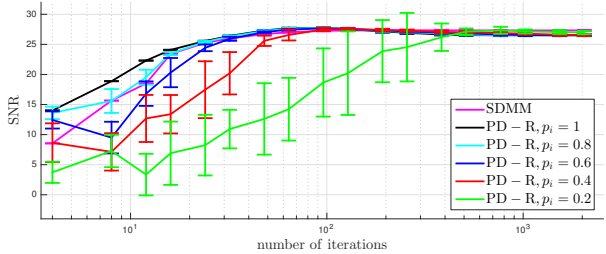}
	\vspace{-2px}
	\caption{The SNR as a function of the number of iterations for \ac{sdmm} and \ac{pd-r} with uniform equal probabilities.}
	\label{fig-m31-10M-eq}
	\vspace{-10px}
\end{figure}

Fig. \ref{fig-m31-10M-eq} contains the \ac{SNR} as a function of the number of iterations for \ac{sdmm} and \ac{pd-r} with equal probabilities $p_i, \forall i$.
For $p_i=1$, when no randomisation is performed, \ac{pd-r} achieves a rate of convergence similar to \ac{sdmm}.
Its computational cost per iteration is much lower even without any randomisation, since \ac{sdmm} computes the solution to a system of equations at each iteration.
If $p_i$ decreases, the average number of nodes needed is reduced as well, to $n = \sum_{i=1}^d p_i$.
In this case, the number of iterations required for convergence increases roughly inversely proportional to $p_i$, reaching a proportionality factor of $2$ at the lowest probability considered.
The cost per iteration for processing the data decreases proportional to the probability.
Thus, if the main computational bottleneck is due to the 
data fidelity term and not to the computation of the Fourier transform, the overall computational cost for reaching convergence does not increase significantly. Here, the increase is up to a factor of $2$ for the lowest probabilities.
The randomised updates offer greater flexibility especially when the number of computing nodes is limited.

\begin{figure}
	\centering
  	\includegraphics[trim={0px 0 0px 0}, clip, width=0.9\linewidth]{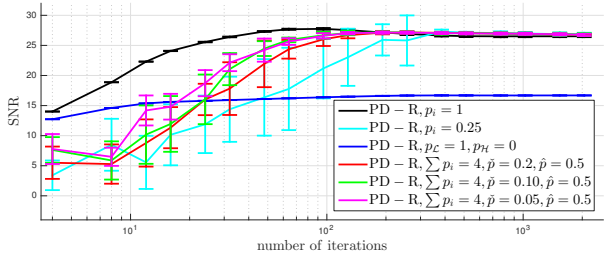}
	\vspace{-2px}
	\caption{The SNR for \ac{pd-r} as a function of the number of iterations.
	On average $4$ blocks are processed per iteration. The algorithm with $p_i=1$ is also included.}
	\label{fig-m31-10M-4Nd}
	\vspace{-10px}
\end{figure}

In Fig. \ref{fig-m31-10M-4Nd} we study the evolution of the \ac{snr} of the randomised algorithm given a limited infrastructure with $n=4$ nodes and $16$ data blocks.
We enforce $\sum_{i=1}^d p_i=4$ such that on average $4$ data blocks are processed on the $4$ computing nodes available.
We compare the adaptive nonuniform probability choice proposed herein with uniform probabilities $p_i=0.25, \forall i$.
We show the results for $\hat{p} = 0.5$. For larger $\hat{p}$ similar results are produced with a slight decrease in convergence speed when $\hat{p}$ approaches $1$.
Different $\check{p}$ are considered.
For comparison, we also include a simulation where we update only the $4$ low frequency blocks with probabilities $p_{\mc{L}}=1$ and ignore the high frequency ones by setting their probabilities $p_{\mc{H}}=0$.
This shows how much the performance degrades if the high frequency data is discarded and motivates the use of randomisation.
A test with no randomisation showcases the fastest achievable convergence speed. This comes at the cost of requiring $16$ data processing nodes.
The proposed adaptive choice for the probabilities substantially increases the rate of convergence.
For convergence it requires only half of the number of iterations needed by the uniform probability test with $p_i=0.25$.
This essentially stabilises the overall total computational cost required for convergence making it roughly the same to that of the algorithm performing no randomisation.
It also lowers the variability of the convergence speed.
Looking at it from a slightly different perspective, the full convergence curve for $\check{p}=0.05$ appears similar to that produced for a uniform probability $p_i=0.4$, the latter requiring on average $n=6.4$ data nodes, a $60\%$ increase, to process the data in a similar time.
Finally, in Fig. \ref{fig-images-and-coverage} we also showcase the reconstructed quality of \ac{pd-r} with $\sum_{i=1}^d p_i=4$, $\check{p}=0.05$ and $\hat{p}=0.5$. The quality is very similar to that achieved by the other tests for PD-R and SDMM.

\section{Conclusions}
\label{sec-conc}

We studied a distributed solver for \ac{ri} imaging that uses randomised updates over the measurement blocks and we proposed a new adaptive strategy for choosing the randomisation probabilities.
The algorithmic structure is highly parallelisable and has a lower computational burden if compared to existing solvers such as \ac{sdmm}.
The algorithm requires little configuration and produces consistently stable results.
The adaptive choice of probabilities is based on the relative approximation error for each data block and is shown to accelerate the convergence speed.
The approach also reduces the variability in the rate of convergence compared to the uniform randomisation.
Our experiments suggest great scalability which is fundamental for the processing of data from next generation telescopes.

\balance
\bibliographystyle{IEEEtran}
\bibliography{abrev,intro,convex-opt,radio-interferometry,sara,signal-proc}

\end{document}